\documentstyle[preprint,aps,amssymb]{revtex}
\input{epsf.tex}
\begin{document}

\title{\bf Wetting transitions of Ne}

\author{M. J. Bojan$^{1,4}$, G. Stan$^2$, S. Curtarolo$^{2,3}$, 
W. A. Steele$^1$, and M. W. Cole$^{2}$ }   
\address{ $^1$Department of Chemistry,
Penn State University,
University Park, PA 16802, USA           \\
$^2$ Department of Physics, Penn State University,
University Park, PA 16802, USA \\
$^3$University of Padua, Physics Department, Padua, Italy \\   
$^4${\em e-mail address: mjb@chem.psu.edu}\\ 
{\em phone: (814)865-2895, fax: (814)863-5319} }

\date{\today}
\maketitle

\baselineskip=22pt

\newpage
\begin{abstract}
We report studies of the wetting behavior of Ne on very weakly attractive
surfaces, carried out with the Grand Canonical Monte Carlo method. The
Ne-Ne interaction was taken to be of Lennard-Jones form, while the
Ne-surface interaction was derived from an ab initio calculation of
Chizmeshya {\em et al}. Nonwetting behavior was found for Li, Rb, and Cs in
the temperature regime explored (i.e., $T < 42$ K). Drying behavior was
manifested in a depleted fluid density near the Cs surface. 
In contrast, for the case
of Mg (a more attractive potential)  a prewetting transition was found
near T= 28 K. This temperature was found to shift slightly when a
corrugated potential was used instead of a uniform potential. The isotherm
shape and the density profiles did not differ qualitatively between these
cases.

\end{abstract}

\newpage
\section{Introduction}

During the last few years there has developed great interest in the
existence and nature of wetting transitions of adsorbed simple gases 
\cite{dietrich}. By ``wetting'' we mean complete wetting, i.e.,
the formation of a film with
thickness that diverges as the pressure approaches saturation.
The problem of wetting has long been interpreted qualitatively in terms of
the relative magnitudes of the cohesive energy of the adsorbate and the
adhesive energy binding the adsorbate to the substrate \cite{pandit}. 
If the latter is small compared to the former, then a simple surface energy argument implies that the gas does not wet the surface at low temperature (T). 
Some 20 years ago it was argued that such a gas will undergo a wetting
transition (of some kind) as T is increased to the critical temperature
\cite{cahn,ebner}. 
The argument has been made both qualitatively, in terms of general
arguments involving entropy and/or surface tension, and quantitatively
using several varieties of model calculations.

Experimental evidence for such transitions was initially slow to appear. In
1992, the first example of a  prewetting transition was found, for the case
of He adsorption on Cs \cite{rutledge,nacher}. 
Prewetting is  a first order  transition manifested by a coverage jump as a 
function of vapor pressure (P). 
This transition had been predicted to occur \cite{cheng1} as a consequence 
of the extremely weak attraction; the well depth $D$ of the He attraction 
to a Cs surface (0.6 meV) is even smaller than the well depth $\epsilon$ 
of the interaction between two He atoms (0.95 meV)! 
These observations were  followed by qualitatively similar evidence (from 
theory and experiment) of prewetting transitions of H$_2$ films on 
Rb and Cs \cite{cheng2,mistura}. 
These findings lead one to ask about  homologous systems, such as Ne films 
on alkali metals. 
There are several simple ways to appraise a given adsorption system 
and estimate its wetting properties. 
If one evaluates the dimensionless ratio $D^*=D/\epsilon$, 
one finds that $D^{*}$ is smaller for Ne on Cs than for any other 
adsorption system in this class; 
see Table ~\ref{table:ratio} \cite{chizmeshya,vidali}. 
This weak relative attraction implies a nonwetting situation at low T.
Further support for this expectation was obtained from a simple model 
\cite{cheng3} which compares the cost (surface tension) of forming a thick 
film with the benefit (the gas-surface attractive interaction). 
This approach gives the following relationship which is satisfied at
the wetting temperature ($T_w$)
\begin{equation}
(\rho_l - \rho_v) I_V \ = \ 2 \gamma
\end{equation}
\begin{equation}
I_V \ = \ - \ \int_{z_{min}}^\infty dz\ V(z)
\end{equation}
where $\rho_l$ and $\rho_v$ are the densities of the adsorbate liquid and
vapor at coexistence, $\gamma$ is the surface tension of the liquid,
and $z_{min}$ is the equilibrium distance of the
potential.
This model predicts that wetting transitions for Ne/Rb and Ne/Cs 
should occur within a few per cent of the Ne critical temperature 
($T_c=44.4\ K$). 
Very recently, experimental support for that prediction was found in 
quartz microbalance measurements on these surfaces. 
The data of Hess {\it et al.} \cite{hess} imply that Ne undergoes a wetting 
transition within 2\% of $T_c$ on Rb and that it undergoes a drying 
transition on Cs. 
The latter is a unique observation, to the best of our knowledge. 
We note in passing that qualitatively similar prewetting transitions have  
been seen recently in Hg adsorption on Mo and sapphire surfaces 
\cite{kozhevnikov,hensel}.

These observations have led us to undertake computer simulation studies of
Ne adsorption on various surfaces \cite{bojan}. 
We have used the Grand Canonical Monte Carlo (gcmc) method, which 
assumes that the behavior of the Ne atoms is classical. 
The accuracy of this assumption for Ne in this temperature range
has been demonstrated recently in detailed calculations of Johnson and
coworkers \cite{wang}. 
We have assumed a Lennard-Jones interatomic interaction between the 
Ne atoms, which should be qualitatively adequate. 
Finally, we have considered two regimes of adsorption interaction. 
One set of calculations pertains to the ultraweakly adsorbing alkali
metal surfaces. Our finding of nonwetting behavior is
qualitatively consistent with the experimental results, within
the limited regime of our study. 
The presence of critical fluctuations prevents us from approaching very 
close to $T_c$ because there the correlation length exceeds the size of 
the periodically replicated unit cell on the surface (2.8 nm). 
The second set of calculations pertains to the case of Ne on much 
more attractive surfaces (e.g. Mg). 
There we find clearly defined prewetting transition behavior. 
That system has not yet been studied experimentally.

Our calculations are not the first simulation studies of the prewetting
phenomenon. 
Previous studies by several groups on a model surface yielded 
prewetting transition behavior for an assumed model potential 
(Ar on CO$_2$) 
\cite{finn,fan,soko}. 
Curiously, previous and subsequent studies, by Nijmeijer, Sikkenk and 
coworkers, did not find 
that transition for a similar system \cite{nijmeijer,sikk,sikk2}. 
Possible reasons for the discrepancy include the role of fluctuations 
and metastability, which are prominent and troublesome in most simulations 
of this problem. Also, while Nijmeijer and coworkers 
used a very large system size in their simulations, 
their choice of the constant volume, constant temperature algorithm
was not as well suited to the study of prewetting as the 
isobaric isothermal simulations used by Monson {\it et al} \cite{finn,fan}.
In our study, 
we are using the gcmc simulation
technique, for which the chemical potential of the system is known. 
Although it
is still possible to obtain metastable states in gcmc, 
we found the problem to be greater at higher temperatures and therefore
were obliged to perform more extensive (longer) simulations in the
metastable regions at high temperatures. 

The outline of this paper is the following. 
Section 2 describes the potentials used as well as our simulation 
methods. Section 3 describes the results found for the ultraweakly 
adsorbing surfaces. Section 4 presents our results for more strongly
adsorbing surfaces including the Ne/Mg case. 
These include analyses for the case of an assumed continuum substrate 
(translationally invariant potential) and a more realistic substrate 
(corrugated potential). 
Section 5 summarizes our results and describes other work and 
predictions which are relevant to this problem.

\section{Model Potentials and Monte Carlo simulations}
The Ne-Ne interaction used in this study is the well-known
Lennard-Jones 12-6 potentials. Therefore, the results
reported here are
relevant to any classical inert gas as well as to a number of simple
molecular gases for which such potentials are appropriate. Consequently,
we can use energies reduced by $k$ (units of K),
remembering that $\epsilon_{Ne-Ne}/k$=33.9 K gives the correct reduced
critical temperature $T^*_c$ = $kT_{cr}/\epsilon_{gg}$=1.31 \cite{smit} for
a Lennard-Jones gas.
The cutoff distance used for
the potential was rather large (5$\sigma$) since it has been noted 
previously \cite{soko} that a small cutoff gives different results for the
fluid properties. For example, in the simulations of Finn and Monson 
\cite{finn}, using
a potential cutoff of 2.5 $\sigma$ gives a critical temperature, 
$T^*_c$ = 1.23.
Although we did not evaluate the critical temperature, 
the condensation pressures of Ne in our simulations agree with the 
vapor pressures as given by Lotfi {\it et al.} \cite{lot}, whose work confirms
the value of $T^*_c$ = 1.31 for a Lennard-Jones fluid.

The gas-solid potential is derived
from a model in which a gas atom interacts with the atoms in the
solid via pairwise Lennard-Jones functions \cite{bruch} and thus
has a somewhat different form from that expected for gases
interacting with a metal surface. Replacing the sums over the solid
atoms in planes parallel to the surface with integrals gives a
gas-solid energy $U_{gs}(z)$ which is:
\begin{equation}
U_{gs}(z)=\epsilon_{1s}\sum_{j=0} \left[ \frac{2}{3}
\left( \frac{\sigma_{gs}}{z+jd}\right)^{10} - \frac{5}{3}
\left( \frac{\sigma_{gs}}{z+jd}\right)^{4} \right] + corr
\end{equation}
where z is the perpendicular distance of the adsorbate atom from the surface,
$\epsilon_{1s}$ is the well-depth, 
the sum has been truncated after $j=3$,
and $corr$ denotes a small correction term for interaction with the
more distant planes.
The relationship between
the well-depth $D$ for the
interaction of an atom with the entire surface and $\epsilon_{1s}$ is
$D \approx 1.2\epsilon_{1s}$ for the interplanar spacing chosen for this
study \cite{steele}.
$\sigma_{gs}$ is the size parameter
of the potential and was taken to be .501 nm; d is the
distance between planes in the solid
and has been set equal to .401 nm  in this work. 
In particular, the relatively large values of $\sigma_{gs}$
will give long-ranged potentials similar to those for the 9-3 functions
used in the theoretical treatments of inert gases on alkali metals
\cite{chizmeshya}.
(See Fig.~\ref{fig:pot} for a comparison of these potentials along with
the potential for the interaction of inert gases with CO$_2$.)
The range of the potentials for atoms over alkali metal surfaces is
particularly large due to the large decay lengths of the surface
electronic charge in these systems \cite{dec}. 

Various well-depths
were chosen for study, ranging from $D =16.8\ K$  to $ 120\ K$
so that the transition from non-wetting
to wetting behavior in these systems could be characterized. 
The studies by Finn and Monson \cite{finn}, Fan and Monson\cite{fan}, 
Soko\l owski and Fischer \cite{soko} correspond to the relative well depth
$D^* = 2.64$. 
The Dutch group (Nijmeijer, Sikkenk, {\it et al.} 
\cite{nijmeijer,sikk,sikk2}) reported simulation results 
covering a wide range of $D^*$ values 
similar to ours (D$^* \approx$ 0.5 to 3.5), 
but they were not able to identify unambiguous
prewetting transitions because their simulations were done in 
the canonical ensemble.

The application of the gcmc algorithm
to the study of adsorption is now a well-known
and extensively utilized technique \cite{AT,FrS,nich}. The present
and previous simulations  of wetting systems determine adsorption isotherms
by evaluating the average number of particles ${\overline N}$ in a computer box
for fixed values of box volume $V$, chemical potential
$\mu$ and temperature $T$. The box has periodic boundary condition
in two dimensions. At least one of the remaining two walls
is the adsorbing surface, with the final wall being either a
second adsorbing surface or a hard wall that serves only
to keep the gas molecules in the box. The second of these options was
used in the present work. Of course, this requires that these two
walls be rather widely separated so the presence of the
second wall has negligible influence on the wetting or adsorption
properties of the system. One method of monitoring this
is to evaluate the local density of the fluid in the system. If
the varying local densities observed in the regions near the
walls are separated by a large region of constant density, one may
reasonably argue that the adsorption is in contact with a phase
having bulk properties. (It is this assumption that underlies
nearly all simulations of physical adsorption.)
Of course there is no assurance
that the phase having constant density is the bulk equilibrium
phase. For the simulations
reported here, these two walls were separated by 7.5 nm;
auxiliary simulations with both larger and smaller wall separations
exhibited no observable dependence of the thermodynamic
properties of the fluid near the wall upon distance
in this range.

The chemical potential of an adsorbed phase is related to
experimentally measurable quantities
by equating it to $\mu$ for the bulk. If the bulk
phase is an ideal gas, one can calculate
the chemical potential from $kT\,ln\,p$. However, this approximation
fails rather badly for the pressures of interest in the present
work, due to deviations from ideality in the bulk gas phase. For
example,  the leading term in the difference $\mu_{gas}-\mu_{id}$
for the real and the ideal gas at the same pressure is
given by:
\begin{equation}
\frac{\mu_{gas}-\mu_{id}}{kT}=-ln[1+Q]+2Q
\end{equation}
where $Q=([4Bp/T+1]^{1/2}-1)/2$, with $B$ equal to the
second virial coefficient of the gas. The thermodynamic
properties of the Lennard-Jones 12-6 fluid have been carefully simulated
by Lotfi {\it et al.} \cite{lot}. Using the values of the vapor pressure
of the bulk phase and the expressions for the non-ideality correction
to the chemical potential given in Lotfi {\it et al.}'s paper, 
one can evaluate
the corrections needed to pass from the chemical potential used
in the gcmc simulation to the pressure in the adsorption cell.
The values given in Table~\ref{table:param} show the magnitudes of these corrections
for the temperatures selected for study and for
pressures equal to the vapor pressure
of the bulk Lennard-Jones liquid \cite{lot}.
The second virial correction to $\mu_{id}$ was found to be
accurate over much of the pressure and
temperature range of the simulations to be reported here, but a
reasonable all-orders expression is also available \cite{lot}
when the first order correction is inadequate.

In the Monte Carlo simulation of a system with variable number of
particles, one has three kinds of move: displacement, creation and
destruction. The relative numbers of these moves must be optimized to
attain equilibrium and give precise average numbers of molecules
with appropriate expenditure of computer resources.
The fractions of trials devoted to particle displacement, to creation
and to destruction amounted to 1/3 each at low densities. As the
number of successful trials decreased with increasing
adsorbate density, the fraction of displacements was decreased to 20\% and the
total number of trials was increased.

Preliminary simulations
showed that the systems studied could be non-wetting
with a nearly first order transition at the bulk vapor pressure
or wetting with an isotherm that exhibited rapid variations
in the number of particles with pressure as the pressure
approached that of the bulk liquid. In such cases, the attainment
of equilibrium required large numbers of
Monte Carlo trials. Thus for systems not close to the steep
regions of the isotherms, the numbers of trials amounted to
$0.5$ to $2\times 10^6$ for the approach to equilibrium and
to $1\times 10^6$ for
evaluation of averages after equilibrium had been attained.
Different initial states for the Monte Carlo chains were taken,
depending upon the expected value of $\overline N$. If  $\overline N$
was small, either an initially empty box or a
configuration from the adjacent isotherm point could be
utilized. However, as the system
approached the condensation region, the optimum initial state
was found to be
a box with a uniform density of particles, equal to roughly
$0.5$ to $0.8$ of that for the bulk liquid; as such a system
evolved toward
equilibrium, a running calculation of the number of particles 
in the
box showed that, aside from fluctuations, it slowly changed
in the direction of the
eventual equilibrium value. Indeed, the criterion
for equilibrium was that this number no longer showed any systematic
variation as the number of trials increased. The technique showed
quite clearly that much larger numbers of trials were needed
to attain equilibrium when the system was in the region of a rapid
change in the number of particles with pressure. Consequently,
the numbers of trials for these points were
$36\times 10^6$ to achieve equilibrium
and $6-12\times 10^6$ to generate  averages after equilibrium had
been reached.

The lateral dimensions $L_x, L_y$ of the box were set at 10
$\sigma_{Ne-Ne}$ or 2.78 nm; consequently, the number of atoms
held in the box when it is full of liquid is $\simeq 1800$ at the lowest
temperature considered.  Because of the relatively large volume of
this box and the weakness of the adsorption, the correction needed
to pass from ${\overline N}$, the total atoms in the box, to
${\overline{N_a}}$, the (thermodynamic excess) 
number of atoms adsorbed, was non-trivial.
One evaluates $N_{gas}$ from the known pressure, the estimated box
volume that is accessible to gas atoms, and the temperature and
the equation of state of the bulk. The accessible box volume is estimated from
the values of $z$ where the local gas density 
decreases to zero, and is necessarily inexact because this decrease
is not infinitely steep. When the adsorption is quite small, as
found for non-wetting systems, the uncertainty in ${\overline{N_a}}$,
the difference
between ${\overline N}$ and $N_{gas}$, relative to ${\overline N}$,
can be quite  significant.

\section{Ultraweak substrates}
Adsorption isotherms and local densities were simulated for the reduced
temperatures listed in Table ~\ref{table:param}. 
For convenience, the corresponding
unreduced temperatures for neon are also listed; they show that
the simulations cover the range from a few degrees below the boiling
point ($27.1\; K$) up to the vicinity of the critical temperature
($44.4\; K$). For the temperatures listed, the well-depth $D$ was set
equal to 16.8, 24, 48, 60, 90, 95, 114, and 120 $K$
in a series of simulations. 
The theoretical well-depths of Ne vary between 24 K (Cs) and 50 K (Li)
for the alkali metals and is 95 K for Mg. The smallest well-depth in the 
series was based on a previous result for Ne on Cs \cite{zaremba}
which is now believed to be too small \cite{chizmeshya}.

Fig.~\ref{fig:D14} shows adsorption isotherms 
calculated for $D = 16.8\; K$.
The discontinuities in all curves at svp indicate that the gas does not wet
this surface at any of the temperatures considered, although the
decrease in the height of the vertical jump as temperature increase
leads one to wonder whether wetting might be present for $T^*$
higher than 1.257 but less than the reduced critical temperature of 1.31.
The surface coverages just preceding the jumps are quite small, but
difficult to specify in terms of layers adsorbed because the monolayer
is  poorly defined in such weakly interacting systems.
A nominal estimate of the monolayer coverage is 
$N_{mono} \sim \rho_{\ell}^{\frac{2}{3}} \cal A \sim $ 60 molecules, based
on the bulk liquid density $\rho_{\ell}$ and surface area $\cal A$.

Note that one can calculate the low coverage parts of these isotherms
from Henry's Law, which states that $N_a = p K_H$, with $K_H$ given
by:
\begin{equation}
K_H=\frac{{\cal A}}{kT} \int [e^{(-U_{gs}(z)/kT)} - 1] dz
\end{equation}
If the lower limit for $z$ is taken to be 1.2 $\sigma_{gs}$ (see Fig.  
~\ref{fig:T1148D14}),
$K_H$ ranges from 0.5 to 0.08 atm$^{-1}$ for
the temperatures listed in Table ~\ref{table:param} and $D=16.8\; K$. 
We see that this expression
predicts extremely small numbers of atoms adsorbed,
at least for the linear
parts of the adsorption isotherms in Fig. ~\ref{fig:D14}.
The analogous
numbers for $D=120\; K$ are 63.0 to 3.3 atm$^{-1}$.

The key point revealed in Fig.~\ref{fig:D14} is that neon does 
not wet a surface
with $D = 16.8\; K$ for all temperatures considered; it is possible
that wetting may take place in the small temperature interval between
the highest $T$ considered here and $T_{cr}$, as found experimentally for Rb.
\cite{hess}. In addition to the isotherms, local densities were
evaluated from the simulations for this system. Thus, the
density is plotted as a function of z, the
distance from the surface, in Fig.~\ref{fig:T1148D14} 
for $T^*=1.148$. The first point to
note is that there is a large range of distance for which the densities
in all cases are constant and equal to the bulk densities. 
Furthermore, when the simulation 
cell is completely filled with liquid, the density profile at the metal
surface is not the same as the density at the hard wall
despite the ultraweak nature of the adsorbing potential.
The second feature is the absence of atoms near both boundary surfaces;
this is not surprising for the hard wall boundary, but the behavior
near the ``adsorbing" surface seems to be the signature of a non-wetting
fluid in a container which is otherwise full of liquid. The experiments
of Hess {\it et al.} \cite{hess} on Cs
have also revealed a region of density depletion near and above
the saturated vapor pressure, in qualitative 
correspondence between this model and the actual Ne/Cs metal system.
Based on the available simulation results, we can suggest that the system
either undergoes a wetting transition close the critical point, 
or given the density profiles, it is possible that the system undergoes 
a drying transition. We are currently trying to simulate this system
in the liquid region to see if a drying transition does indeed occur.

Simulated isotherms for D = 24 K (corresponding to Ne adsorbed on
Cs) were also obtained at the same
temperatures as those shown for D = 16.8 K and are not 
significantly different.
Results for D = 48 K (Ne on Li)  are shown in Fig.~\ref{fig:Liisot}.
Here the system appears to be nonwetting for all temperatures studied
despite the prediction by Chizmeshya (based on Eq. 1) 
that Ne should wet Li above 33 K.
It is possible that the prewetting regime is so close to saturation
that our simulations cannot discern it. Alternately, 
metastability in this region may be preventing us from obtaining 
unambiguous results. 

\section{Stronger substrates}
The gas-solid interaction in the Ne/Mg system is approximated by a 
well depth $D$=95 K with a wetting temperature of 26 K
predicted by the simple surface tension model given in
equation 1
\cite{chizmeshya}. In this case, the transition is 
convenient for both experiment
and computer simulations (the wetting temperature being far from 
the critical temperature). 
Therefore, we explored a range of temperatures around the predicted 
T$_w$ and found  prewetting transitions for temperatures in the
predicted range.
Fig.~\ref{fig:D80pressure} shows the simulated adsorption isotherms 
for the temperatures considered.
At T=21 K, the adsorption seems to be small up to saturation, 
while for temperatures between 22 and 29 K
prewetting transitions are observed. 
Thus we estimate $T_w=22 \pm 1 K$ from our simulation data for Mg. This
compares favorably with a prediction of 22 K from Eq. 1, using the
potential assumed in the simulation. Hence we conclude that the heuristic
Eq. 1 works well in this case \cite{note}.
It is important to note one fact; the potential used
in the simulations is seen in Fig. 1 to have a wider attractive bowl than
the ab initio potential of Chizmeshya. 
Thus we conclude that the best estimate of the wetting temperature
on Mg is actually 26 K (as reported in Chizmeshya's paper \cite{chizmeshya}).

The second important temperature is the prewetting critical temperature. We
estimate its value to be 31 K since that is where the isotherms begin to
become continuous; this is a provisional value for two reasons. One is that
the simulations cannot accurately treat a critical transition, due to size
effects. The other is that the potential difference discussed for $T_w$ would
imply that 31 K is probably an underestimate relative to what would be
predicted with the Chizmeshya potential. Setting aside these concerns, we
may compare our results for the reduced prewetting critical temperature,
$T^*_{pw}=0.88$, with the values reported by others for the Ar/CO$_2$ system. 
See Table~\ref{table:Tw}.
Because the $D/\epsilon$ ratio for the Ne/Mg and Ar/CO$_2$ systems are 
nearly the same, we attribute the differences in the wetting behavior
of the two systems
to the difference in the width of the potential wells. The width used for
the Ne/Mg study is $w = 0.229$ nm while $w$ for the  potentials used
in the various Ar/CO$_2$ studies is 0.178 nm.
Here ``width'' means the full width at half minimum of the attractive part of
the potential. Increasing the value of $w$ increases the integral, I$_v$
(equation 1). This increase in the left hand side of the
equation implies a decrease in the wetting temperature (and the prewetting
critical temperature).

The prewetting behavior seen in the Ne/Mg isotherms can also be
characterized by its density profiles.
In Fig.~\ref{fig:T30D80density}, densities are 
given for T=28 K. 
The curves shown correspond to density profiles for points on the isotherm
immediately before and immediately after the transition at P = 1.07atm.
Before the transition, there is little adsorption as the thin film is favored.
However, there is an enhanced density at the surface relative to the
gas phase density in the region of attractive potential. 
This is in sharp contrast to
the density profile for the $D$=16.8 K case where the density of the
fluid in most of the simulation box is equal to the gas phase density (See
Fig.~\ref{fig:T1148D14}). After the transition, three distinct peaks can be seen in the 
density profiles, indicating that the prewetting behavior corresponds to
a transition from less than a monolayer to about 2.5 to 3 layers;
the total coverage jump 20 
 is a factor of 15. 
The separation between the peaks close to the surface
is 0.93 $\sigma_{gg}$ for the the first two and 0.98 $\sigma_{gg}$
between the second and third peaks. The spacing between
close-packed (111) planes of bulk Ne is 
0.261 nm = 0.94 $\sigma_{gg}$.

In order to see the dependence of the adsorption and wetting properties
upon gas-solid interaction strength, we have simulated adsorption
for several other values of
D while holding the temperature fixed. 
Fig.~\ref{fig:D4i} shows isotherms at  a high temperature (39 K)
for some values of D considered while Fig.~\ref{fig:T30pressure} gives
similar results for a lower temperature (T = 28 K). 
At the high temperature, the curves for 90 and 120 K show the
smooth increase of coverage with pressure as one approaches the
saturated liquid, but that for 60 K is sharply different. At first
glance, the apparent discontinuity in slope  appears to be the
signal of a non-wetting surface, but the fact that the jump in coverage
occurs at a pressure slightly less than that of the saturated liquid
actually indicates a prewetting transition. Furthermore,
the local densities that are plotted in Fig.~\ref{fig:T1148D50} 
for the 60 K system show
peaks close to the adsorbing surface  which are due to
monolayer formation. As noted above and in other
simulation studies of wetting \cite{sikk,adam},
these peaks will generally be absent for the condensate formed in non-wetting
systems.
One also observes in Fig.~\ref{fig:T1148D50} 
that the density fluctuations in the monolayer
region are smaller than those farther away, at z $>$5 $\sigma$. 
This behavior is
consistent with the following argument. In a slowly varying system,
the number fluctuation $\Delta$ N within a very small volume will satisfy the
theorem known to apply to a uniform system:

\begin{equation}
\langle(\Delta N)^2\rangle = kT \frac{dN}{d\mu}
\end{equation}
where the derivative is that of a uniform system at the local density, with
a chemical potential value shifted from that of the nonuniform system by
the net external potential. This is an application of so-called ``local
thermodynamics''.  For qualitative purposes, and often quantitative ones,
one may apply this approach to the highly nonuniform vicinity of the
monolayer. Here one finds a much smaller density fluctuation for the (quite
intuitive) reason that the derivative in that region is much smaller than
for larger z; the monolayer density profile does not change much with the
addition of particles at larger z.

For the low temperature study
the pressures for the prewetting transition
are shifted to lower pressures relative to the bulk vapor pressure.
Thus, isotherms at 28 K clearly exhibit the complete range of wetting
behavior as the well-depth is varied from 95 K to 114 K. 
At $D$=60 K (not shown), the system is nonwetting, 
while at $D$=114 K the adsorption
is continuous, with prewetting transitions occurring
for systems of intermediate strengths. Unfortunately, 
the fluctuations in the transition regions are larger at this
temperature and for well-depths below 95 K
we have not obtained results in the transition region that
were sufficiently precise to give us a good prediction of the transition 
pressure. However, the fact that there are large fluctuations is a good
indicator that a first order transition is taking place. In the continuous
and non-wetting regions at this temperature, no such problems occurred. 

Finally, we look to see what happens if we modify the potential to take
into account the corrugation in the surface. We constructed a potential
by adding Ne-Mg pair interactions, assuming a Mg
surface with a simple cubic structure (lattice parameter of .401 nm)
and chose a well-depth for this potential such that the 
well-depth of the lateral
average of the corrugated potential is identical to that of
the smooth potential.
The potentials are compared in 
Fig.~\ref{fig:D80potential} and
the resulting isotherms are shown in Fig.~\ref{fig:D80T30corrug}. The 
periodicity causes a shift in the pressure at
which the prewetting transition occurs and therefore brings about a lowering
of the wetting temperature. This is presumably due to the deeper potentials at 
the adsorption sites, but other factors (such as the lateral spacing of
adsorbent atoms) could play a role in this phenomenon \cite{other}. 
The density distributions which correspond to coverages immediately 
before and immediately after the prewetting transition are given in
Fig.~\ref{fig:T30D80density} for the flat surface and in 
Fig.~\ref{fig:D80T30dens_corrug} for the corrugated potential model.
These plots are very similar, indicating that the nature of the transition 
has remained the same despite the increased corrugation.
The principal difference is a $\sim$60\% higher film density just below
the transition in the corrugated surface case. 

\section{Summary}
In this paper we have examined the nature of the wetting transition of Ne
on various weakly adsorbing surfaces. For the case of the most inert
surface, Cs, we find negligible adsorption of Ne throughout the temperature
range  explored. In fact, for pressures (slightly) above saturated vapor
pressure, we find evidence of a dry region near the surface. This behavior
is qualitatively similar to what was found experimentally by Hess,
Sabatini, and Chan. 
In the case of somewhat more attractive surfaces, e.g.
Mg, we find evidence of a prewetting regime of temperature. The adsorption
there is characterized by a jump in coverage at a specific pressure near,
but below, saturation. The coverage changes by a factor of order five to
fifteen at this transition. This finding is consistent with both experiments
(inert gases on alkali metal surfaces and Hg on sapphire) and 
with general trends established in the simulations
of Monson and coworkers.

While there is every indication that this prewetting transition behavior is
generic on weakly attractive surfaces, more extensive study is needed to
assess the criteria for the prewetting phenomenon. One useful guide to its
occurrence is the simple model, Eq. 1, which predicts the wetting
temperature from the potential and the bulk fluid's properties. This was
found to work surprisingly well for the case of Mg, but for the Li 
surface, the results are ambiguous. The simple
dependence contained in that relationship implies that one may invert the
wetting transition temperature obtained experimentally in order to deduce
an integral property of the potential. This capability can be particularly
useful in circumstances where the gas-surface attraction is weak, because
then the Henry's law regime of low coverage may be inaccessible or
difficult to utilize as a means of analyzing the potential.

We have found qualitatively similar results for the cases of a smooth
adsorption potential and a corrugated potential (similar to what
might occur on a Mg surface). It is not surprising that the effect of the
periodic potential is small at the transition pressure because most of the
adsorbate lies in the region of negligible corrugation and the 
amplitude of the Fourier components of the potential are $< kT$ at the 
potential minimum.

The prewetting transition has thus far been investigated in relatively few
adsorption systems. We hope that our conceptually straightforward findings
encourage experimentalists to extend their studies to other adsorption
systems which are weak-binding systems. Some of these other cases have
rough estimates of wetting temperatures presented  by Chizmeshya {\it et al}.

In future publications we would like to explore two problems of further
interest. In one, we will report results of a study of the effects of
roughness on the wetting transition. We hope at a later date to evaluate
the temperature regime close to the critical point where wetting
transitions have been found for Ne on Rb and Cs.

{\acknowledgements}  We are most grateful to Victor Bakaev for sharing his
expertise in Monte Carlo simulations and for helpful discussions.
We are also grateful to George Hess, Moses Chan and Ian
McDonald for stimulating conversations. This research was supported
by the National Science Foundation.
We are especially grateful to {\it Fondazione Aldo Ing. Gini} for 
the generous support of Ing. S. Curtarolo.

\newpage

\begin{table}  [h]
\center
\caption {Potential ratios}
The ratio $D^*$ of the adsorption potential well depth D to a
nominal pair potential well depth $\epsilon$ (in parentheses, in Kelvin) is
tabulated for simple gases on various surfaces.  Pair well depths
are taken from reference \cite{watts}, except for the case
of H$_2$, which is taken from Cheng  {\em et al.} \cite{cheng3} and Ne, 
which is fit to the bulk critical temperature (44.4K). Adsorption
well depths are from Chizmeshya {\em et al.} \cite{chizmeshya} except
Au \cite{cheng3} and CO$_2$ \cite{ebner}.
\vspace*{0.2 in}

\begin{tabular}{|c|c|c|c|c|}
                  & He       &Ne      &H$_{2}$         &Ar  \\
                  &(10.22)   &(33.9)  &(34.3)        &(119.8)\\
        \hline
         Au        & 8.5      & 8.3    & 13            & 8.2 \\
         Mg        & 3.4      & 2.8    & 5.5           & 3.5\\
         Li        & 1.7      & 1.5    & 2.8           & 2.0 \\
        Rb        & 0.72     & 0.71   & 1.1           & 1.1  \\
        Cs        & 0.68     & 0.70   & 1.1           & 1.1  \\
        CO$_2$    & ---      & ---    & ---           & 2.64     \\
\end{tabular}
\label{table:ratio}
        \end{table}

\begin{table}  [h]
\center
\caption{Parameters of the  simulations }
(Here, $\sigma$ denotes liquid-vapor equilibrium)
\vspace*{0.2 in}
\begin{tabular} {|c|c|c|c|c|} 
$T^{*}$ & $T\; (K)$  & p$_{\sigma}$ (atm) &
$e^{(\mu_{real}-\mu_{id})_{\sigma}/kT}$ & $e^{(\mu_{real}-\mu_{id})
_{\sigma}/kT}$ \\
                 & (L.J. neon) & (L.J. neon) & (all orders) &
(second virial)  \\
\hline
0.656 &  22  &  0.1424  &   0.9889  &   0.989 \\
0.738 &  25  &  0.4828  &   0.9725  &   0.973 \\
0.812 &  28  &  1.2249  &   0.9478  &   0.948 \\
0.902 &  31  &  2.5788  &   0.9154  &   0.915 \\
1.011 &  34  &  5.7238  &   0.862   &   0.862     \\
1.093 &  37  &  9.4306  &   0.817   &   0.815     \\
1.148 &  39  & 12.7238  &   0.784   &   0.777      \\
1.230 &  42  & 19.0158  &   0.732   &   0.714      \\
1.257 &  43  & 21.5500  &   0.716   &   0.692      \\
\end{tabular}
\label{table:param}
\end{table}

\begin{table}  [h]
\caption{Wetting temperatures for Ne/Mg and Ar/CO$_2$ studies.}
\vspace*{0.2 in}
\begin{tabular} {||c|c|c||} 
Source & $T^*_w$ & $T^*_{pw}$ \\ \hline 
Ebner and Saam \cite{ebner} & 0.77 & 0.92 \\
Evans and Tarazona \cite{Tara} & 0.957 & 0.988 \\
Meister and Kroll \cite{Meis} & 0.90 &   \\
Finn and Monson \cite{finn} & 0.84 & 0.94 \\
Soko\l owski and Fischer \cite{soko} & 0.975 $\pm$ 0.025 & \\
This work & 0.65 & 0.88 $\pm$ 0.03 \\
\end{tabular}
\label{table:Tw}
\end{table}

\newpage
\begin{figure}
\caption {Summed 10-4 potential (---) used in this study is
compared to the integrated 9-3 potential for the interaction of Ne
with alkali and alkaline earth metals
(--- ---) and the 9-3 potential for the Ar/CO$_2$
interaction ($\cdots$). For the 10-4 potential, z$_{min}$
is 0.497 nm, while for the integrated potentials, z$_{min}$ is the distance
above the jellium edge, and is 0.307 nm  for the Ne/metal 
potential, and 0.320 nm  for the Ar/CO$_2$ potential.}
\label{fig:pot}
\end{figure}

\begin{figure}
\caption { Adsorption isotherms for a Lennard-Jones gas (Ne)
on a model Rb surface with gas-solid interaction well-depth
$D= 16.8$ K. 
Shown are isotherms for T=43 K
($\circ$ and ---), T=42K ($\ast$ and --- -- ---),
T=41 K ($\triangle$ and --- $\cdot$ ---), T=39 K ($\Box$ and --- ---), 
T=37 K ($\blacktriangle$ and -- --), T=34 K ($\bullet$ and ---) and  
T=32 K ($\blacksquare$ and -- $\cdot$ --).}
\label{fig:D14}
\end{figure}

\begin{figure}
\caption {Local densities for the gas in the computer box
for the system of Fig.  2
at $T^*=1.148$. The area of this
box is .773 nm$^2$ and the density gives the total number of
atoms per unit volume in the box as a
function of distance from the surface. 
For each curve,
the number of adsorbed atoms (= total atoms minus gas atoms in a 
box of volume V at a pressure p) in the box
is denoted by $N_a$. Pressures are 13.5
atm at $N_a/\cal A$=11.1 nm$^{-2}$ (---)
13.9 atm at $N_a/\cal A$= 137 nm$^{-2}$ ($\cdots$) and 16.5 atm at 
$N_a/\cal A$=183 nm$^{-2}$ (--- $\cdot$ ---). Bulk liquid and
gas densities at the adsorption pressures for each curve are shown by
the horizontal lines. The simulation value of the
vapor pressure of neon at T*=1.148 is given in Table I.}
\label{fig:T1148D14}
\end{figure}

\begin{figure}[ht]
\caption{Adsorption isotherms for a Lennard-Jones gas (Ne) 
on a model Li surface
with gas-solid interaction well-depth 48 K. The temperatures are 43 K
($\times$ and -- --), 42 K ($\circ$ and ---), 39 K ($\ast$ and
--- -- ---), 37 K ($\triangle$ and --- $\cdot$ ---), 34 K ($\Box$ and 
--- ---), 32 K ($\blacktriangle$ and -- --), 28 K ($\bullet$ and ---) and 25 K ($\blacksquare$ and -- $\cdot$ --). 
The vertical lines represent
the saturated vapor pressure at the respective temperatures.}
\label{fig:Liisot}
\end{figure}

\begin{figure}[ht]
\caption {Adsorption isotherms for Ne on a model Mg surface with gas-solid
interaction well-depth $D = 95 K$. The corresponding temperatures are 31 K
($\Box$ and --- ---), 29 K ($\blacktriangle$ and -- --), 28 K
($\bullet$ and ---), 27 K ($\blacksquare$ and --- $\cdot$ ---),
25 K ($\circ$ and -- $\cdot$ --), 23K ($\ast$ and --- ---), 22 K
($\triangle$ and --- $\cdot$ ---), and 21K ($\times$ and --- --- ). 
The vertical lines represent the
saturated vapor pressure at the respective temperatures.}
\label{fig:D80pressure}
\end{figure}

\begin{figure}[ht]
\caption{Local densities as a function of the reduced distance from the
surface for the gas in the model Ne/Mg system ($D = 95 K$) at T=28 K. 
Curves shown are for coverages of 22. atoms/nm$^3$ at P = 1.04 atm (---)
and 303. atoms/nm$^3$ at
P = 1.05 atm ($\cdots$).
The bulk liquid density (--- ---) and
the gas density (-- $\cdot$ --) are represented for comparison.}
\label{fig:T30D80density}
\end{figure}

\begin{figure}
\caption {Adsorption isotherms for a Lennard-Jones gas (Ne) at
$T^*=1.148$ on several surfaces with the
interaction well-depths D=16.8 K ($\blacksquare$ and -- $\cdot$ --),
D=60K ($\bullet$ and ---), D=90 K ($\blacktriangle$ and -- --) 
and D=120 K ($\Box$ and --- ---). The prewetting jump for D=60 K
goes from N$_a/\cal A$ = 17.9 molecules/nm$^{2}$ to N$_a/\cal A$ = 102.7
molecules/nm$^{2}$.} 
\label{fig:D4i}
\end{figure}

\begin{figure}[ht]
\caption{Adsorption isotherms at T=28 K for model systems with gas-solid
interaction strengths slightly different from the model Ne/Mg system.
Shown are isotherms corresponding to well-depths of
95 K ($\blacksquare$ and ---, corresponding 
to Ne/Mg), 102 K ($\bullet$ and -- --), 108 K ($\Box$ and --- ---), 
114 K ($\circ$ and -- $\cdot$ --).}
\label{fig:T30pressure}
\end{figure}

\begin{figure}
\caption {Same as Fig. 3, but for an interaction well-depth
of 60 K. Pressures are 13.0 atm at $N_a/\cal A$=14.5 nm$^{-2}$,
13.5 atm at $N_a/\cal A$= 92.8 nm$^{-2}$, 
13.9 atm at $N_a/\cal A$=152 nm$^{-2}$, and 
14.2 atm at $N_a/\cal A$=170 nm$^{-2}$.}   
\label{fig:T1148D50}
\end{figure}

\begin{figure}[ht]
\caption{Reduced gas-solid adsorption potentials as a function of reduced
distance from the surface for a continuum flat surface ($\blacksquare$,
modeling Ne/Mg), and for a corrugated surface with a square lattice
(lattice constant a=.401 nm). In this last case, the adatom is placed
vertically above three sites: S(---), SP (--- ---), and A (-- $\cdot$ --)
whose location is indicated in the inset.}
\label{fig:D80potential} 
\end{figure}

\begin{figure}[ht]
\caption{Adsorption isotherms at T= 28 K for the Ne/Mg model system in the
case of a continuum flat surface 
($\blacksquare$ and -- --) and a corrugated surface
($\Box$ and $\cdots$).}
\label{fig:D80T30corrug}
\end{figure}

\begin{figure}[ht]
\caption{Same as Fig. 6 but for the corrugated potential. The two
isotherm points that were considered are for coverages of 34. 
atoms/nm$^3$ at P = 0.92 atm (---) and 289. atoms/nm$^3$ at 
P =0.94  atm (-- --). }

\label{fig:D80T30dens_corrug}
\end{figure}

\newpage
\begin{figure}[ht]
\epsfysize=5.in \epsfbox{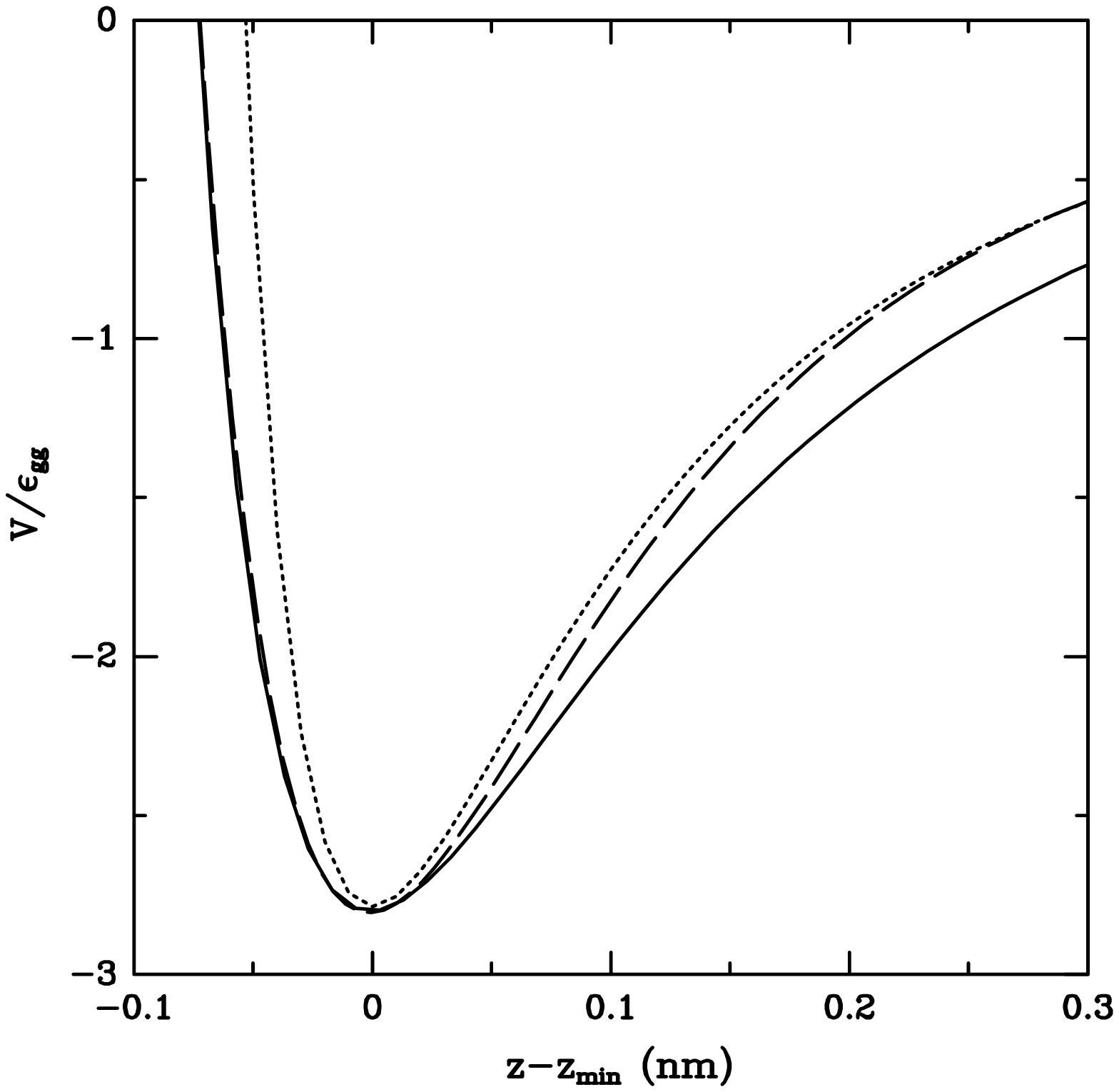}
\end{figure}
\vspace{6cm}
\begin{center}
{\bf FIG. 1}
\end{center}

\newpage
\begin{figure}[ht]
\epsfysize=5.in \epsfbox{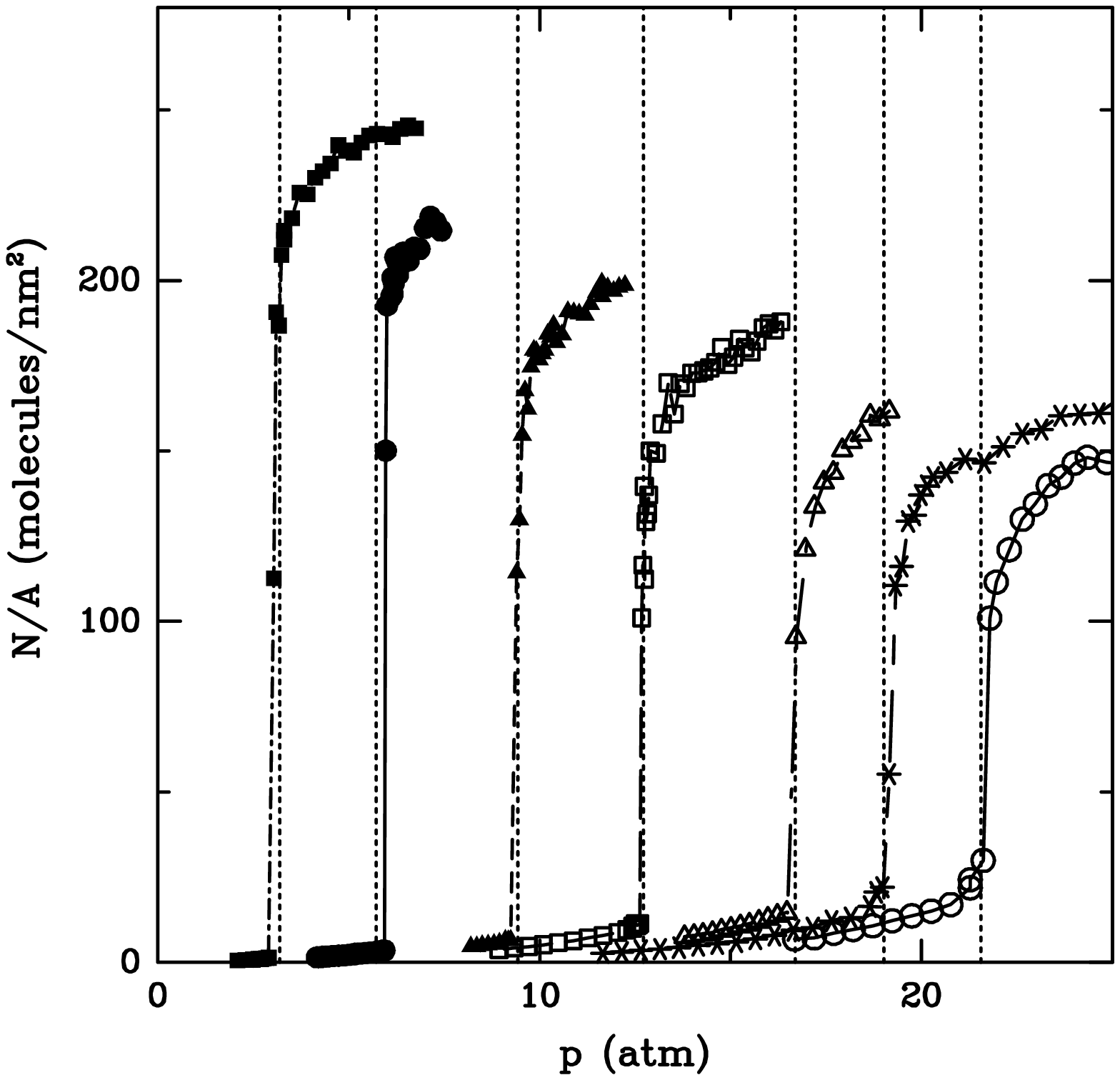}
\end{figure}
\vspace{6cm}
\begin{center}
{\bf FIG. 2}
\end{center}

\newpage
\begin{figure}[ht]
\epsfysize=5.in \epsfbox{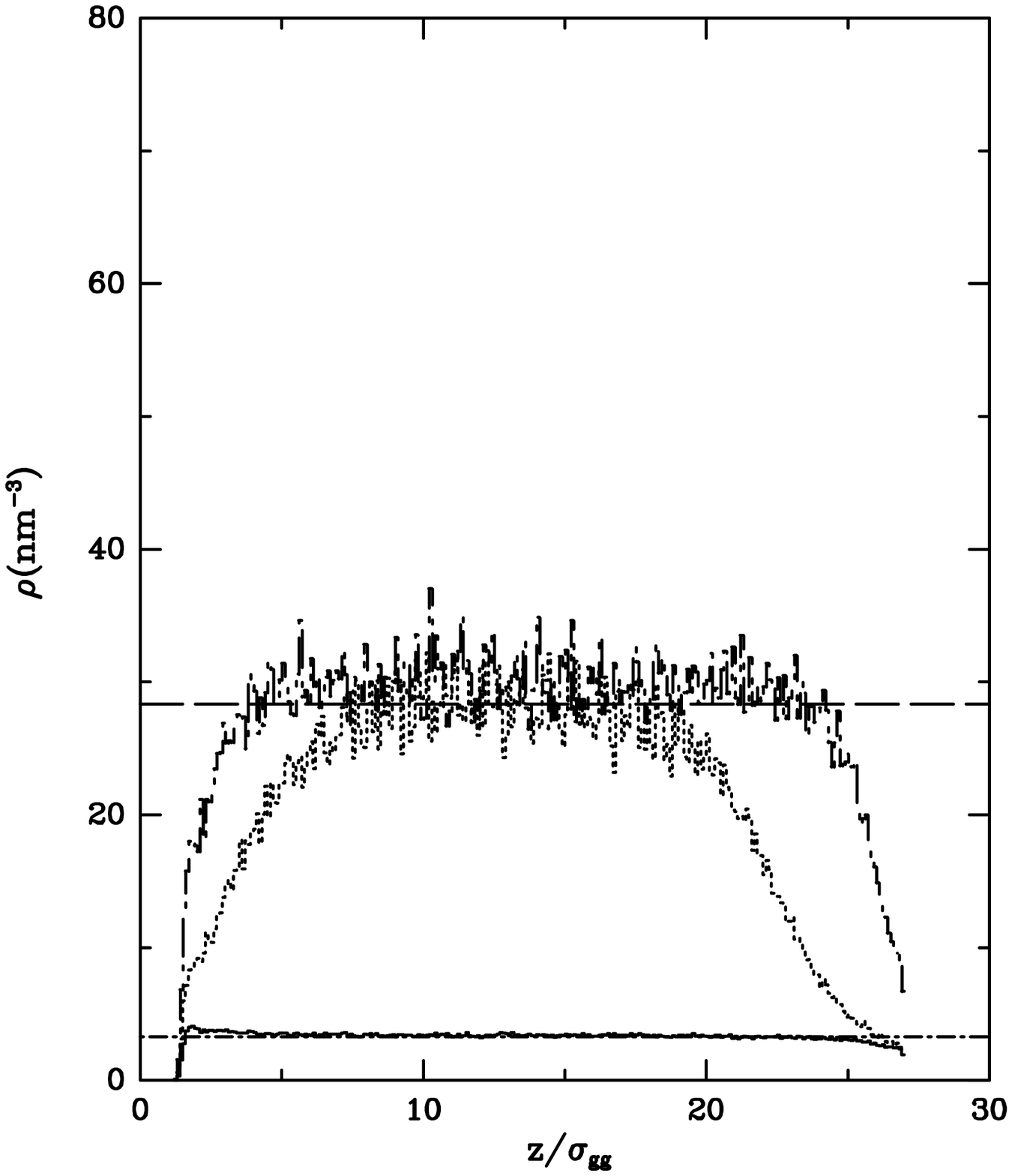}
\end{figure}
\vspace{8.5cm}
\begin{center}
{\bf FIG. 3}
\end{center}

\newpage
\begin{figure}[ht]
\epsfysize=5.in \epsfbox{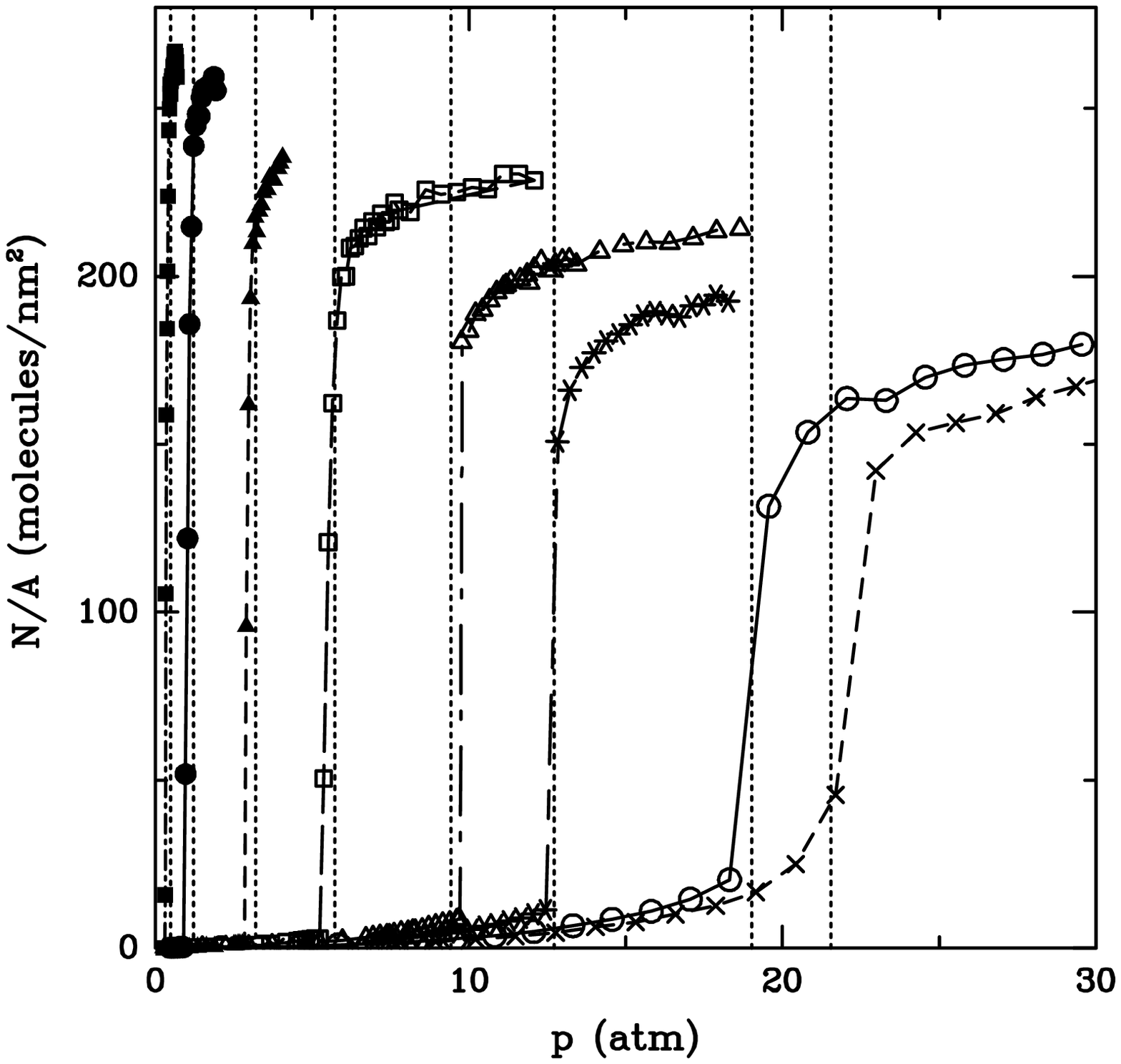}
\end{figure}
\vspace{6cm}
\begin{center}
{\bf FIG. 4}
\end{center}

\newpage
\begin{figure}[ht]
\epsfysize=5.in \epsfbox{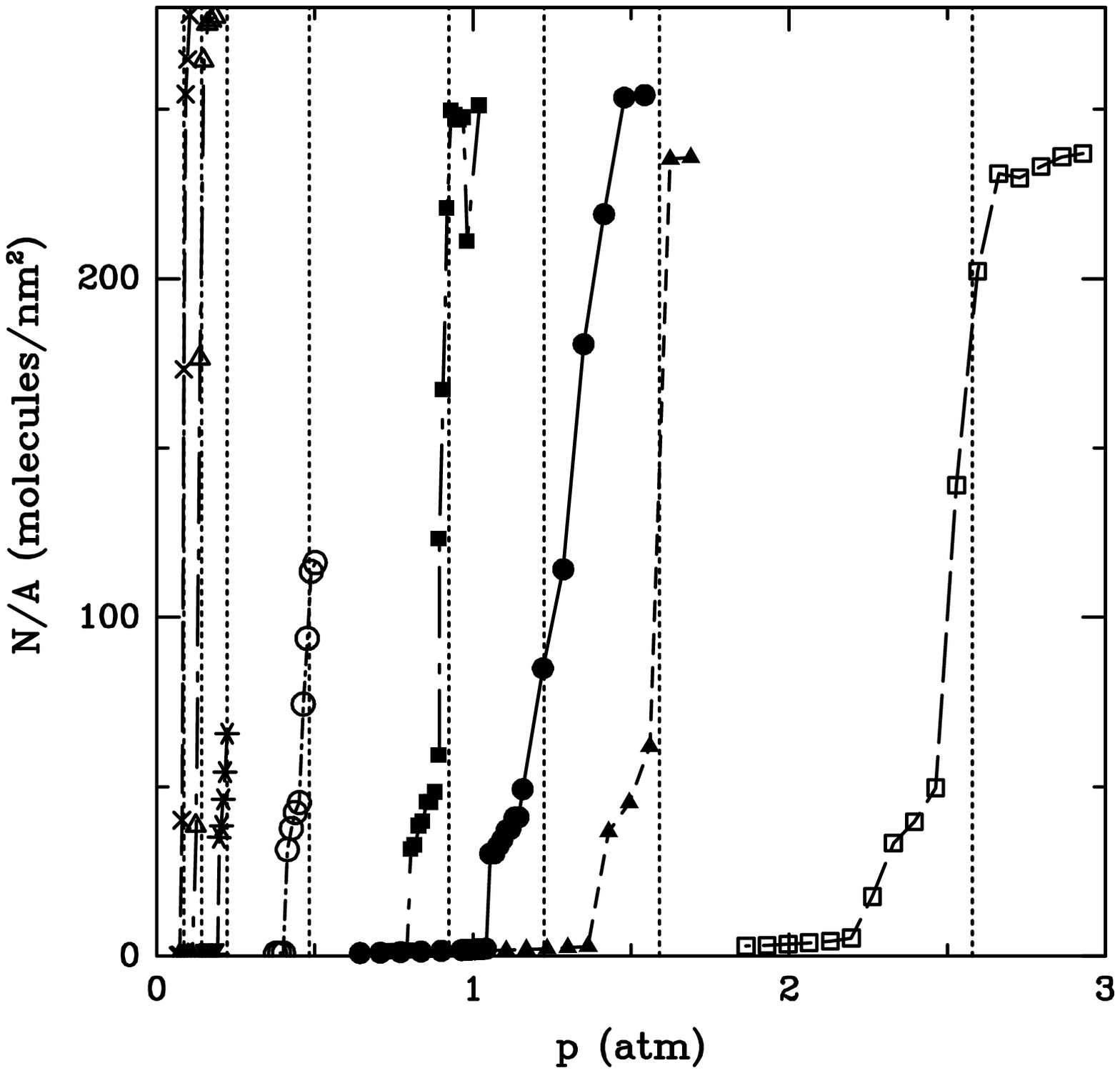}
\end{figure}
\vspace{6cm}
\begin{center}
{\bf FIG. 5}
\end{center}

\newpage
\begin{figure}[ht]
\epsfysize=5.in \epsfbox{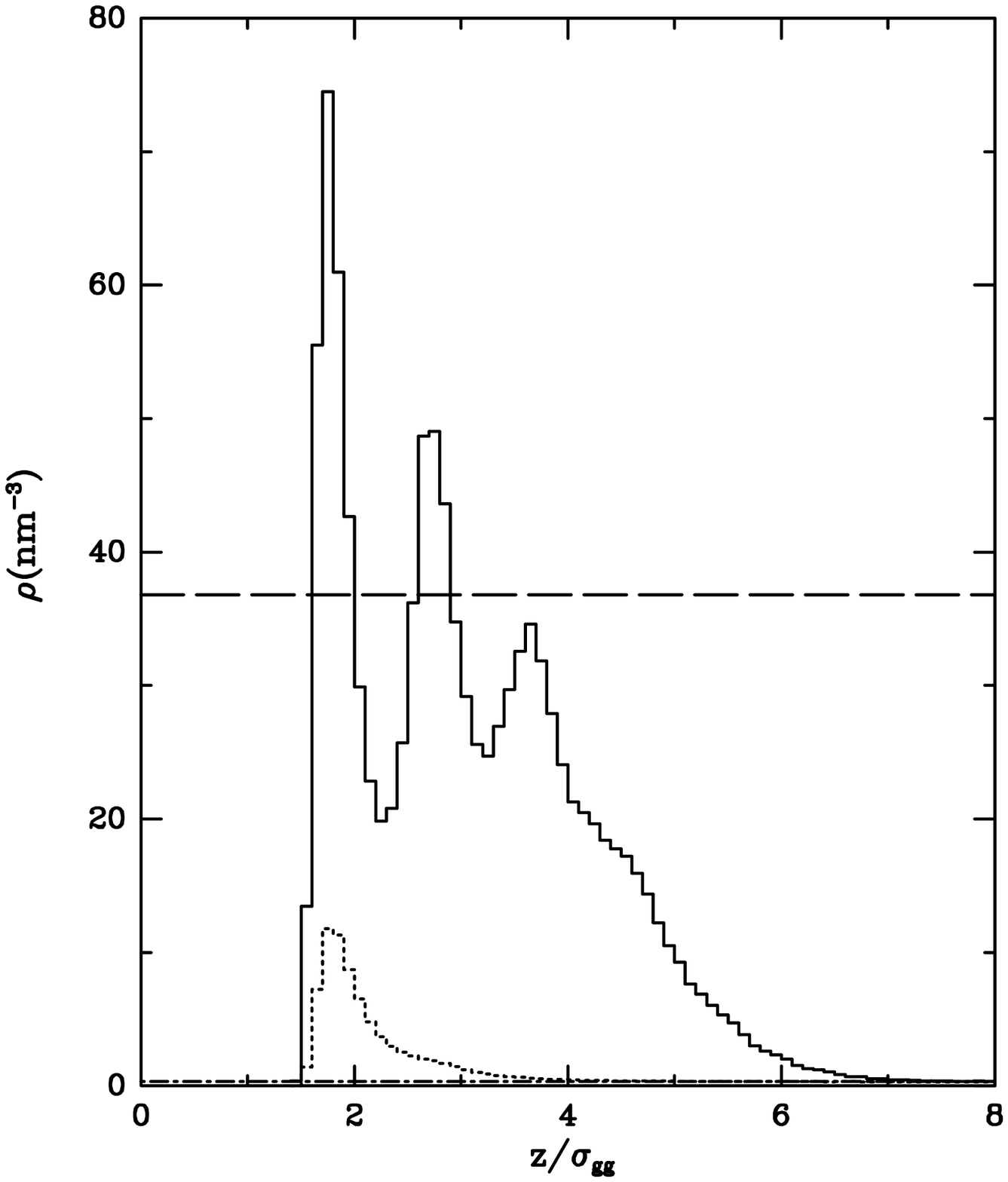}
\end{figure}
\vspace{8.5cm}
\begin{center}
{\bf FIG. 6}
\end{center}

\newpage
\begin{figure}[ht]
\epsfysize=5.in \epsfbox{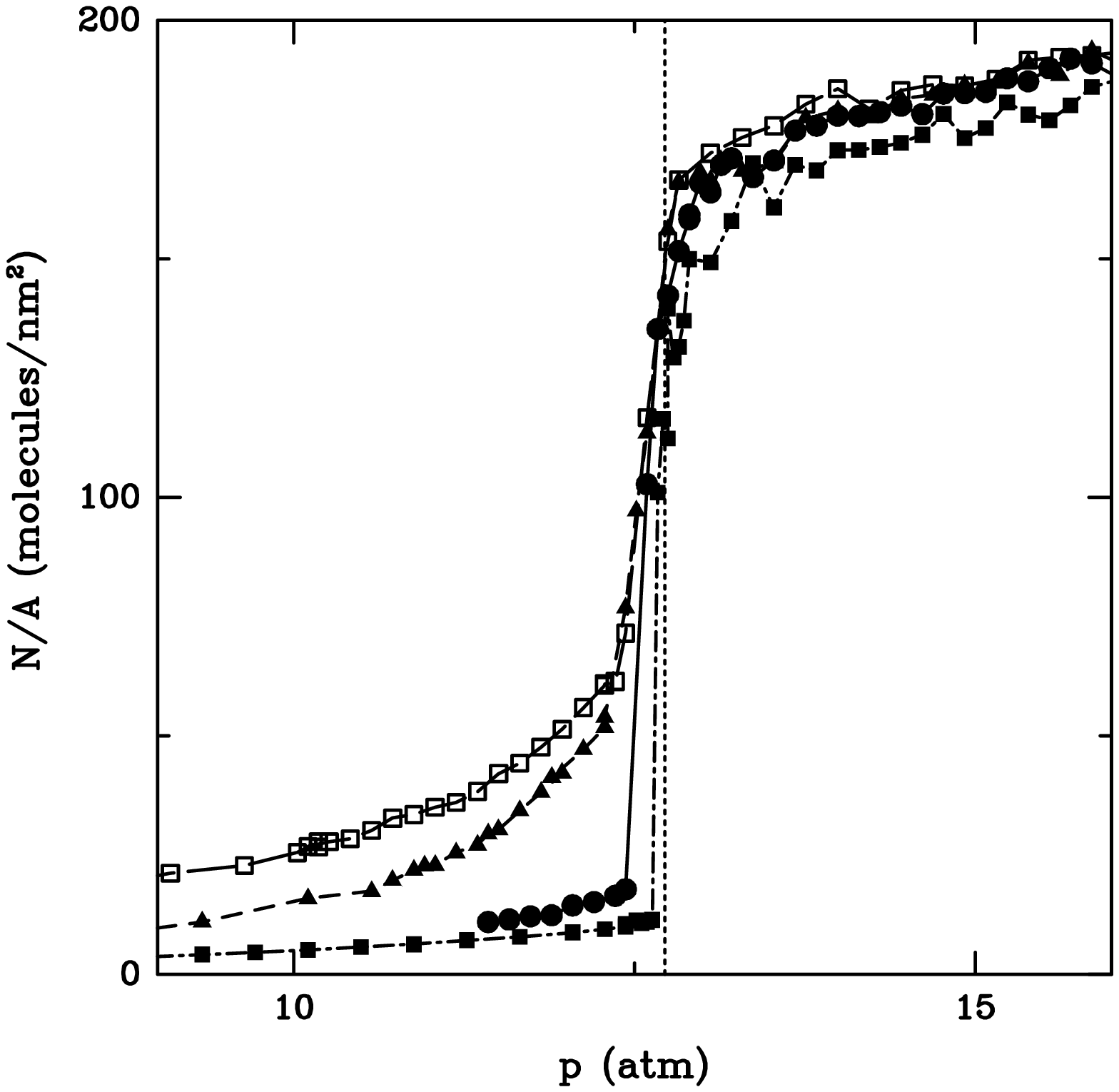}
\end{figure}
\vspace{6cm}
\begin{center}
{\bf FIG. 7}
\end{center}

\newpage
\begin{figure}[ht]
\epsfysize=5.in \epsfbox{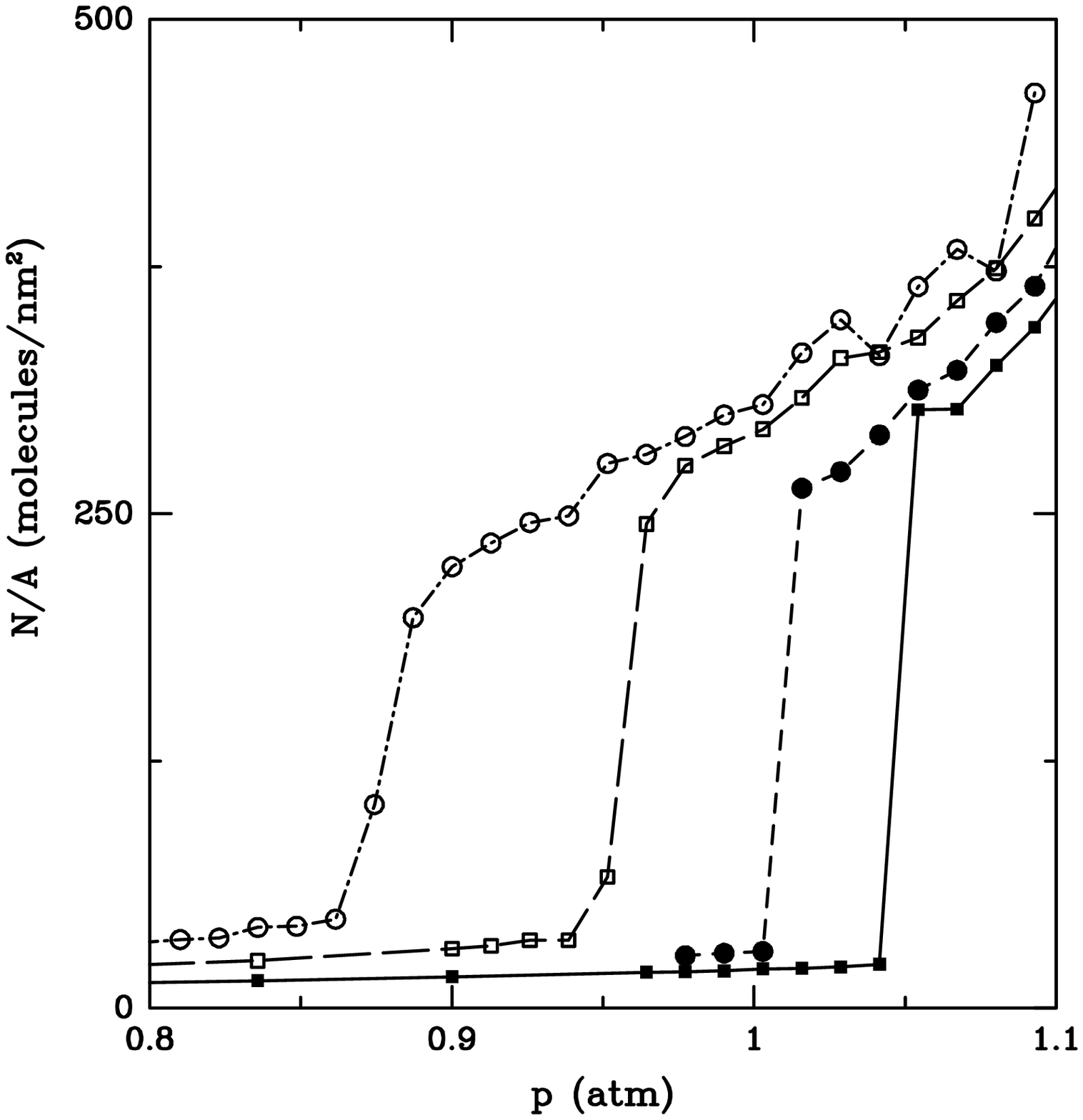}
\end{figure}
\vspace{7cm}
\begin{center}
{\bf FIG. 8}
\end{center}

\newpage
\begin{figure}[ht]
\epsfysize=5.in \epsfbox{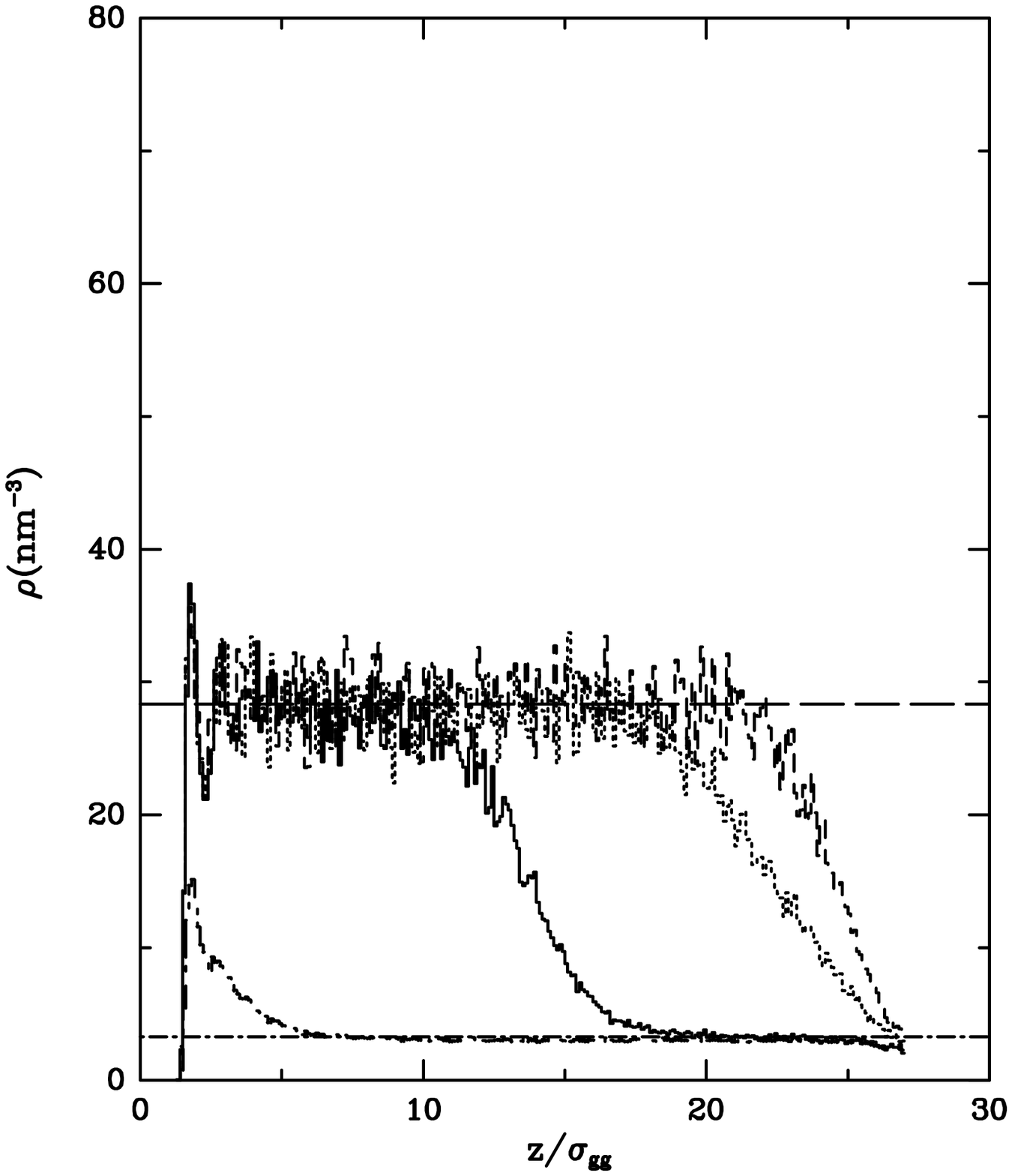}
\end{figure}
\vspace{8.5cm}
\begin{center}
{\bf FIG. 9}
\end{center}

\newpage
\begin{figure}[ht]
\epsfysize=5.in \epsfbox{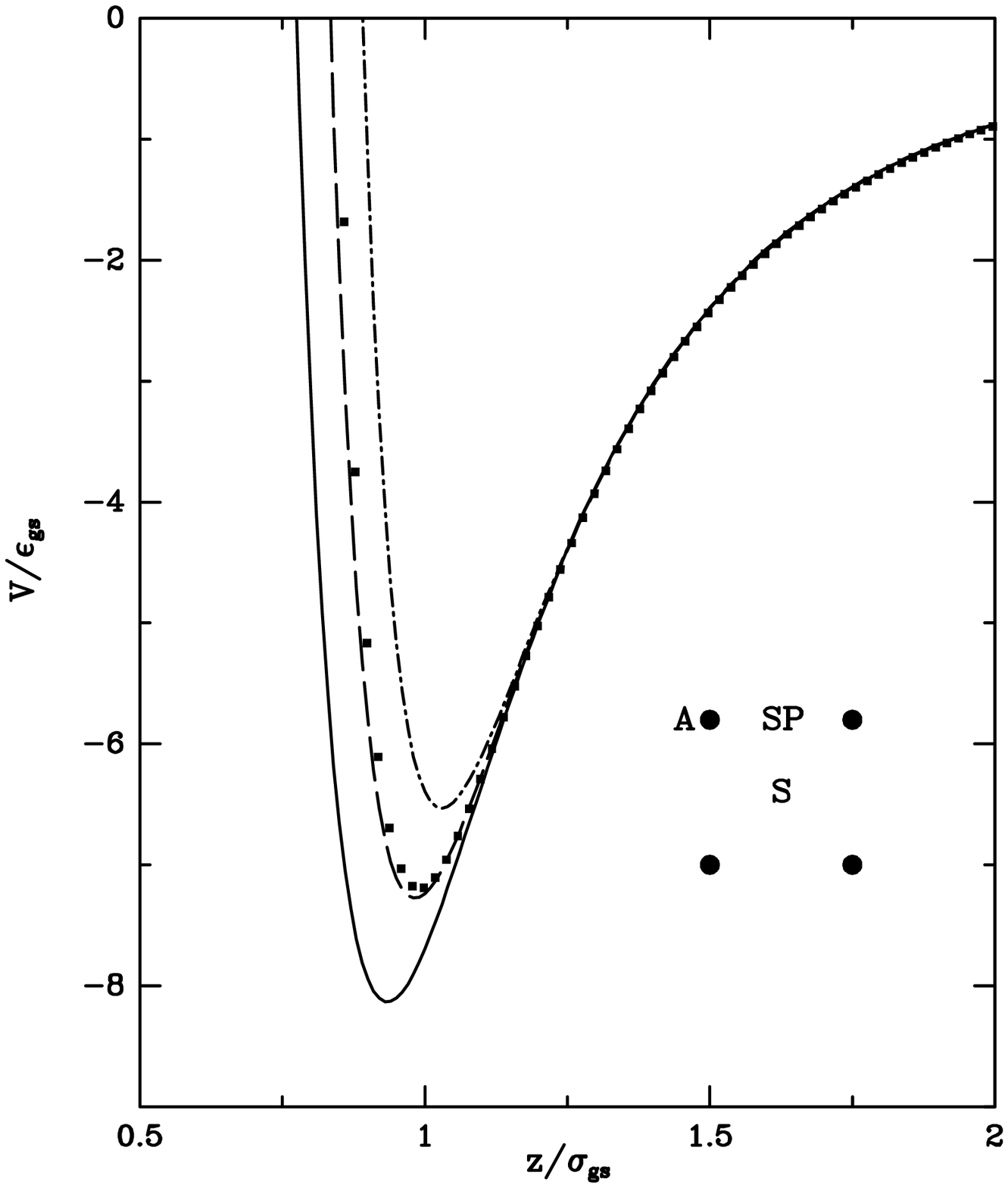}
\end{figure}
\vspace{8.5cm}
\begin{center}
{\bf FIG. 10}
\end{center}

\newpage
\begin{figure}[ht]
\epsfysize=5.in \epsfbox{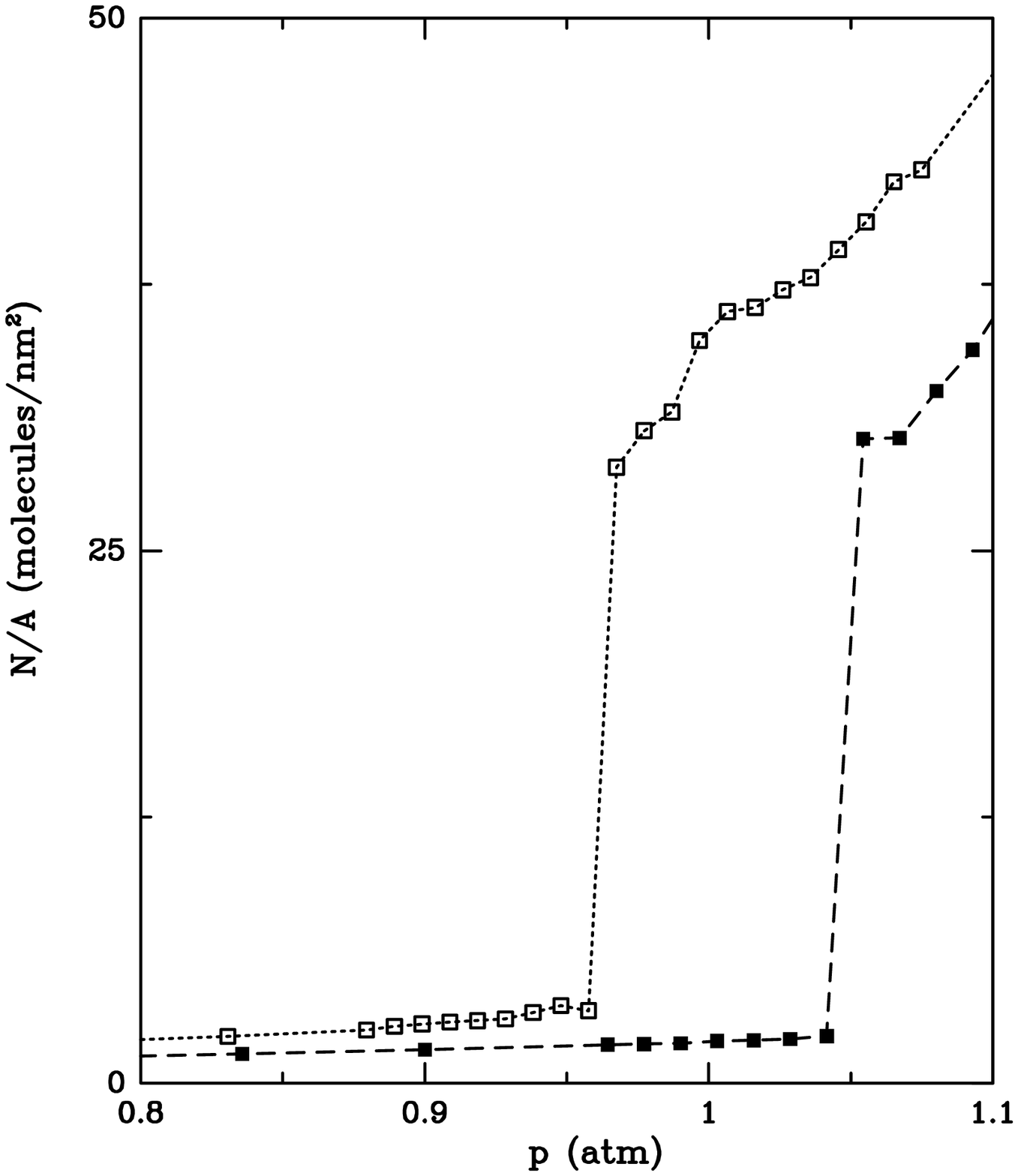}
\end{figure}
\vspace{8.5cm}
\begin{center}
{\bf FIG. 11}
\end{center}

\newpage
\begin{figure}[ht]
\epsfysize=5.in \epsfbox{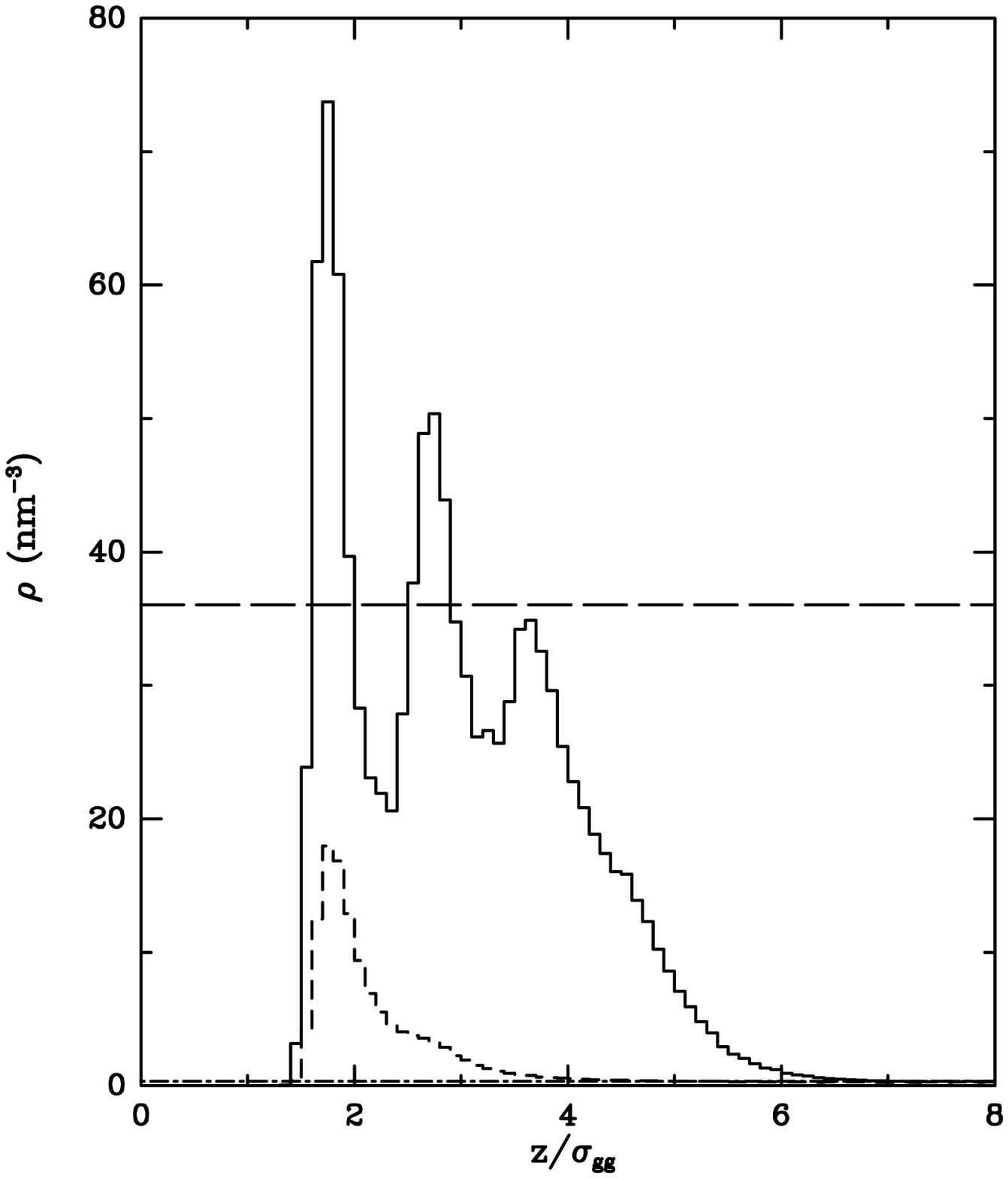}
\end{figure}
\vspace{8.5cm}
\begin{center}
{\bf FIG. 12}
\end{center}


\begin{thebibliography}{50}
\bibitem{dec} The decay length is $2\pi \sqrt{8Wm}/h$, where W and m
are the electronic work functions and mass.
\bibitem{other} We are currently working on this problem.
\end{thebibliography}
\end{document}